\documentclass[twocolumn,aps,prb,superscriptaddress]{revtex4}
\usepackage{ae} 
\usepackage[T1]{fontenc}
\usepackage[ansinew]{inputenc}
\usepackage{amsmath}
\usepackage{graphicx}
\usepackage{color}
\usepackage[colorlinks]{hyperref}
\usepackage{setspace}

\begin{document}

\date{\today}
\title{Thermoelectrical manipulation of nano-magnets}
\author{A. M. Kadigrobov}
\affiliation{Department of Physics, University of Gothenburg, SE-412
96 G{\" o}teborg, Sweden} \affiliation{Theoretische Physik III,
Ruhr-Universit\"{a}t Bochum, D-44801 Bochum, Germany}
\author{S. Andersson}\affiliation{Nanostructure Physics, Royal Institute
of Technology, SE-106 91 Stockholm, Sweden}
\author{D. Radi\'{c}}
\affiliation{Department of Physics, University of Gothenburg, SE-412
96 G{\" o}teborg, Sweden} \affiliation{Department of Physics,
Faculty of Science, University of Zagreb, 1001 Zagreb, Croatia}
\author{R. I. Shekhter}
\affiliation{Department of Physics, University of Gothenburg, SE-412
96 G{\" o}teborg, Sweden}
\author{M. Jonson}
\affiliation{Department of Physics, University of Gothenburg, SE-412
96 G{\" o}teborg, Sweden} \affiliation{School of Engineering and
Physical Sciences, Heriot-Watt University, Edinburgh EH14 4AS,
Scotland, UK}\affiliation{ Division of Quantum Phases
and Devices, School of Physics, Konkuk University, Seoul 143-701, Korea}
\author{V. Korenivski}\affiliation{Nanostructure Physics, Royal Institute
of Technology, SE-106 91 Stockholm, Sweden}
\date{\today}

\begin{abstract}
We investigate the interplay between the thermodynamic properties and spin-dependent transport in a mesoscopic device based on a magnetic multilayer (F/f/F), in which two strongly ferromagnetic layers (F) are exchange-coupled through a weakly ferromagnetic spacer (f) with the Curie temperature in the vicinity of room temperature. We show theoretically that the Joule heating produced by the spin-dependent current
allows a spin-thermo-electronic control of the ferromagnetic-to-paramagnetic (f/N) transition in the spacer and, thereby, of the relative orientation of the outer F-layers in the device (spin-thermo-electric manipulation of nanomagnets). Supporting experimental evidence of such thermally controlled switching from parallel to antiparallel magnetization orientations in F/f(N)/F sandwiches is presented.
Furthermore, we show theoretically that local Joule heating due to a high concentration of current in a magnetic point contact or a nanopillar can be used to reversibly drive the weakly ferromagnetic spacer through its Curie point and thereby exchange couple and decouple the two strongly ferromagnetic F-layers. For the devices designed to have an antiparallel ground state above the Curie point of the spacer, the associated spin-thermionic parallel-to-antiparallel switching causes magneto-resistance oscillations whose frequency can be controlled by proper biasing from essentially DC to GHz. We discuss in detail an experimental realization of a device that can operate as a thermo-magneto-resistive switch or oscillator.
\end{abstract}

\maketitle

\section{Introduction} The problem of how to manipulate magnetic
states on the nanometer scale is central to applied
magneto-electronics.  The torque effect \cite{Slonczewski,Berger},
which is based on the exchange interaction between spin-polarized electrons injected into a ferromagnet and its magnetization,
is one of the key phenomena leading to current-induced magnetic
switching. Current-induced precession and switching of the
orientation of magnetic moments due to this effect have been observed
in many experiments
\cite{Tsoi1,Tsoi2,Myers,Katine,Kiselev,Rippard,Beech,Prejbeanu,Jianguo,Dieny}.

Current-induced switching is, however, limited by the necessity to
work with high current densities. A natural solution to this
problem is to use electrical point contacts (PCs). Here the
current density is high only near the PC, where it can
reach\cite{Yanson,Versluijs} values  $\sim 10^9$~A/cm$^2$. Since
almost all the voltage drop occurs over the PC the characteristic
energy transferred to the electronic system is comparable to the
exchange energy in magnetic materials if the bias voltage $V_{bias}\sim 0.1$ V, which is easily reached in experiments. At the same
time the energy transfer leads to local heating of the PC region,
where the local temperature can be accurately controlled by the
bias voltage.
%
%

Electrical manipulation of nanomagnetic conductors by such
controlled Joule heating of a PC is a new principle for
current-induced magnetic switching. In this paper we discuss one
possible implementation of this principle by considering a
thermoelectrical magnetic switching effect. The effect is caused
by a non-linear interaction between spin-dependent electron
transport and the magnetic sub-system of the conductor due to the
Joule heating effect. We predict that a magnetic PC with a
particular design can provide both voltage-controlled fast
switching and smooth changes of the magnetization direction in
nanometer-size regions of the magnetic material. We also predict
temporal oscillations of the magnetization direction (accompanied
by electrical oscillations) under an applied DC voltage. These
phenomena are potentially useful for microelectronic applications
such as memory devices and voltage controlled oscillators.

\section{ Equilibrium magnetization distribution \label{equilibrium}}
\begin{figure}
 \centerline{\includegraphics[width=0.7\columnwidth]{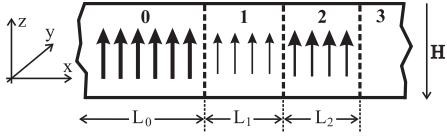}}
  \caption{ Orientation of the magnetic moments in
  a  stack of three ferromagnetic layers
(0, 1, 2) coupled to a non-magnetic conductor (3);  the right arrow
indicates the presence of a magnetic field $H$, which is
antiparallel to the stack magnetization.}
 \label{noflip}
 \end{figure}

The system under consideration has three ferromagnetic layers
coupled to a non-magnetic conductor as sketched in
Fig.~\ref{noflip}. We assume that the Curie temperature $T_c^{(1)}$
of region 1 is lower than the Curie temperatures $T_c^{(0,2)}$ of
regions 0 and 2; in region 2 there is a magnetic field directed
opposite to the magnetization of the region, which can be an
external field, the fringing field from layer 0, or a combination of
the two. We require this magnetostatic field  to be weak enough so
that at low temperatures $T$ the magnetization of layer 2 is kept
parallel to the magnetization of layer 0 due to the exchange
interaction between them via region 1 (we assume the magnetization
direction of layer 0 to be fixed). In the absence of an external
field and if the temperature is
above the Curie point, $T> T_c^{(1)}$, the spacer of
 the proposed F/f(N)/F tri-layer is similar to the antiparallel spin-flop `free layers'
widely used in memory device applications\cite{Worledge}.

\begin{figure}
 \centerline{\includegraphics[width=0.7\columnwidth]{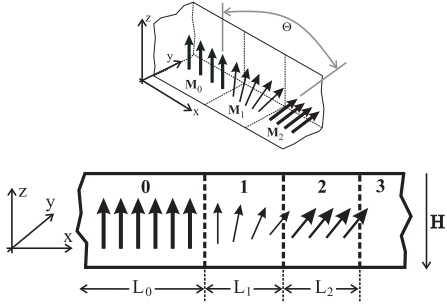}}
\vspace{0cm} \caption{ Sketch of the spatial dependence of the
orientation of the magnetic moments in the stack of
Fig.~\ref{noflip} at a temperature $T$ higher than the temperature
$T_c^{\rm(or)}$, at which the parallel orientation becomes unstable,
but lower than the Curie temperature $T_c^{(1)}$ of layer (1).
 }
 \label{changeorient}
\end{figure}

As  $T$  approaches $T_c^{(1)}$  from below the magnetic moment of
layer 1 decreases and the exchange coupling between layers 0 and 2
weakens. This results in an inhomogeneous  distribution of the
stack magnetization, where the
distribution that
 minimizes the free energy of the system is given by Euler's equation
(see, e.g., Ref. \onlinecite{Landau}):
\begin{equation}
\frac{d}{d x}\left(\alpha (x)M^2(x)\frac{d\theta}{dx}
\right)-\frac{\beta}{2}M^2
\sin{2\theta}+\frac{HM}{2}\sin{\theta}=0\,.\label{inhom}
\end{equation}
Here the $x$-axis is perpendicular to the layer planes of the stack,
the $z$-axis is directed along the
magnetization direction in region 0; the magnetization direction depends only on $x$
and  the vector ${\bf M}$ rotates  in-plane (that is in the $yz$-plane) \cite{Landau};
$\theta (x)$ is
the angle between the  magnetic moment $\vec M (x)$ at point $x$ and
the $z$-axis (in the $yz$-plane) and $M(x)=|\vec M(x)|$.  In the
case under consideration $\alpha (x)= \alpha_1$,
$\beta(x)=\beta_{1}$ for $0 \leq x \leq L_1$ and $\alpha (x)=
\alpha_2$, $\beta(x)=\beta_{2}$ for $L_1 < x \leq L_2$; here
$\alpha_{1,2} \sim I_{1,2}/a M_{1,2}^2$, where $a$ is the lattice
spacing, $I_{1,2}\sim k_BT_c^{(1,2)}$ and $M_{1,2}$ are the exchange
energies and magnetic moments of regions 1 and 2; $\beta_{1}$
($\beta_{2}$) is a dimensionless measure of the anisotropy energy of
region 1 (region 2); $k_B$ is the Boltzmann constant. Below we
assume the lengths $L_{1,2}$ of regions 1 and 2 to be shorter than
the domain wall lengths in these regions $l_{1,2} =
\sqrt{\alpha_{1,2}/\beta_{1,2}}$.

In order to find the magnetization distribution inside the stack one
may solve Eq.~(\ref{inhom}) in regions 1 and 2 to get $\theta_{1}(x)
$ and $\theta_{2}(x) $, respectively, and then match these solutions
at the magnetization interface $x=L_1$. Integrating
Eq.~(\ref{inhom}) with respect to $x$ in the limits $L_1-\delta \leq
x \leq L_1 +\delta, \;\; \delta \rightarrow 0 $ one gets the
matching condition as follows:
%
$$
 \alpha_2 M_2^2\frac { d \theta_2(x)}{dx}\Bigr|_{x=L_1}=\alpha_1
M_1^2\frac { d \theta_1(x)}{dx}\Bigr|_{x=L_1};
$$
\begin{equation}
\theta_2(L_1)=\theta_1(L_1).
 \label{matchingcond}
\end{equation}

The boundary condition at the ferromagnetic interface $x=0$ between
layers 0 and 1  follows from the requirement that the direction of
the magnetization in layer 0 is fixed  along the $z$-axis (i.e.,
$\theta(x) =0$ in this layer):
\begin{equation}
\theta_1(0)=0.
 \label{01boundary}
\end{equation}

At the "free" end of the ferromagnetic sample the boundary
condition for the magnetization $\vec{M}(x)$   is $d \vec{M}(x)/dx =
0$ (see, e.g. Ref.~\onlinecite{Akhiezer}), so that
\begin{equation}
\frac{d \theta_2(x)}{dx}\bigr|_{x=L_1+L_2}=0\,.
 \label{freebound}
\end{equation}

 Solving Eq.~(\ref{inhom}) in regions 1 and 2 under the
 assumption $L_{1,2} \ll l_{1,2}$  and with the boundary conditions
(\ref{matchingcond}) - (\ref{freebound})
 one finds the magnetization in
region 1 to be inhomogeneous,
\begin{equation}
\theta_1(x)= \Theta(L_1)\frac{x}{L_1} + {\cal O}(\frac{L_1}{l_1}); \hspace{0.25cm}  0\leq x \leq L_1 \,,
 \label{layer1solution}
\end{equation}
while  due to the boundary condition (\ref{freebound})  the magnetic
moments in region 2 are approximately parallel, to within
corrections of order $\alpha_1 M_1^2(T)L_2/\alpha_2 M_2^2(T)L_1\ll 1
$, i.e.

\begin{equation}
\theta_2(x)=\theta_2(L_1+L_2)-\frac{H \sin\theta_2(L_1+L_2)}{8\alpha_2
M_2} (L_1 +L_2-x)^2
\label{theta2(x)}\,
\end{equation}
where $L_1 \leq x\leq L_1+L_2$. Using the above boundary conditions one finds that $\theta_2(L_1)\approx \theta_2(L_1+L_2) \equiv \Theta $
is determined by the equation
\begin{eqnarray}\label{theta2}
\Theta=D(H,T)\sin{\Theta}, \hspace{1.0 cm} T < T_c^{(1)} \nonumber \\
\Theta=\pm \pi, \hspace{1.0 cm} T \geq T_c^{(1)}
\end{eqnarray}
where
\begin{equation}
D(H,T)=\frac{L_1 L_2H M_2(T)}{4 \alpha_1 M_1^2(T)}\;\;
 \label{D}\,.
\end{equation}
In Eq.~(\ref{D}) $M_1(T)=M_1^{(0)}\sqrt{(T_c^{(1)}-T)/T_c^{(1)}}$
and $M_2(T)$ are the magnetic moments of region 1 and 2,
respectively; the parameter $D(H,T)$ is the ratio between the
magnetic energy and the energy of the stack volume for the
inhomogeneous distribution of the magnetization. As the second term inside the brackets in
Eq.~(\ref{theta2(x)}) is negligibly small, the magnetization tilt
angle $\Theta$ in region 2 becomes independent of position and is
simply given as a function of $H$ and $T$ by Eq.~(\ref{theta2}).

By inspection of Eq.~(\ref{theta2}) one finds that it has either one
or several roots in the interval $-\pi \leq \Theta \leq \pi$
depending on the value of the parameter $D(H,T)$.

At low temperatures the exchange energy prevails, the parameter
$D(H,T) <1$ and Eq.~(\ref{theta2}) has only one root,
$\Theta=0$. Hence a parallel orientation of all magnetic moments
in the stack is thermodynamically stable. However, at
temperature $T_c^{\rm(or)} < T_c^{(1)}$ for which
$D(T_c^{\rm(or)},H)=1$,
two new roots $\Theta =\pm |\theta_{\rm
min}| \neq 0$ appear. The parallel magnetization corresponding to
$\Theta=0$ is now unstable \cite{phase} and the direction of the
magnetization in region 2 tilts as indicated in
Fig.~\ref{changeorient}. Using Eq.(\ref{D}) one finds the {\bf critical} temperature of
this orientation transition to be equal to
\begin{equation}
T_c^{\rm(or)}= T_c^{(1)}\left(1- \frac{\delta T}{T_c^{(1)}} \right), \hspace{0.2cm}
 \frac{\delta T}{T_c^{(1)}}=\frac{L_1 L_2H M_2}{4 \alpha_1 M_1^2(0)}\equiv D_0 \label{Torient}\,.
\end{equation}
The tilt increases further with $T$ until
at $T = T_c^{(1)}$ the
 exchange coupling between layers 0 and 2 vanishes and their
 magnetic moments become antiparallel.


\section*{Thermally assisted exchange \emph{decoupling} in F/f/F multilayers}

To demonstrate the properties of the tri-layer material system proposed above we have
suitably alloyed Ni and Cu to obtain a spacer with a $T_{c}$ just
higher than room temperature (RT). The alloying was done by
co-sputtering Ni and Cu at room temperature and base pressure
$10^{-8}$ torr on to a 90x10 mm long Si substrate in such a way as to obtain a variation in the concentration of Ni and Cu along the substrate.
By cutting the substrate into smaller samples along the
compositional gradient, a series of samples were obtained, each
having a different Curie temperature.

\begin{figure}[htbp]
\centering \includegraphics[width=0.9\columnwidth]{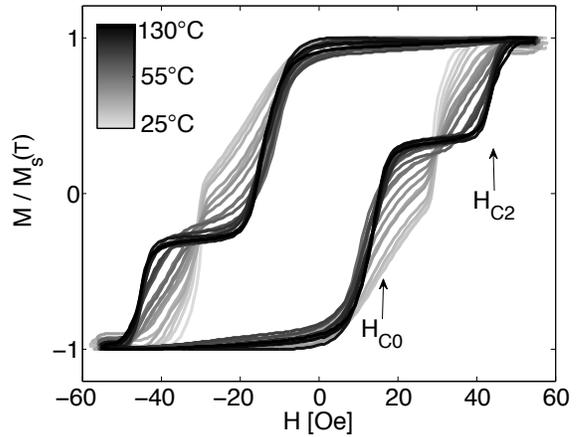}
\caption{Magnetization loop for a sample of SiO$_{2}$/Cu
90/Ni$_{80}$Fe$_{20}$ 8/Co$_{90}$Fe$_{10}$ 2/Ni$_{70}$Cu$_{30}$
30/Co$_{90}$Fe$_{10}$ 5/Ta 10 {[}nm] as the temperature is varied
from $25^{\circ}$C to $130^{\circ}$C. H$_{\mathrm{C}0}$ and
H$_{\mathrm{C}2}$ are the coercive fields of the bottom and top
magnetic layers, respectively.}
\label{fig:MH_vs_T}
\end{figure}

One of the multi-layer compositions chosen was NiFe 8/CoFe 2/NiCu 30/ CoFe 5 $[\rm nm ]$
where the NiFe layer is used to lower the coercive field of the
bottom layer (H$_{\mathrm{C}0}$) in order to separate it from the
switching of the top layer (H$_{\mathrm{C}2}$). A magnetometer
equipped with a sample heater was used to measure the magnetization
loop as the temperature was varied between $25^{\circ}$C and
$130^{\circ}$C. The results for a Ni concentration of $\sim$70\%
are shown in Fig.~\ref{fig:MH_vs_T}. The strongly ferromagnetic
outer layers are essentially exchange-\emph{decoupled} at
$T>100^{\circ}$C (F/Paramagnetic/F state), as evidenced by the two
distinct magnetization transitions at approximately 15 and 45 Oe in
Fig.~\ref{fig:MH_vs_T}. As the temperature is reduced to RT, the
switching field of the soft layer increases and the originally sharp
$M$-$H$ transition becomes significantly skewed. This confirms the
theoretical result, expressed by Eqs.~(\ref{layer1solution}) and
(\ref{theta2(x)}) for $\Theta(H,T)$, that the magnetic state of the
sandwich is of the spring-ferromagnet type \cite{spring_mag}. The
lowering of temperature leads at the same time to a lower switching
field of the magnetically hard layer, which is due to the stronger
effective magnetic torque on the top layer in the coupled F0/f/F2
state. This thermally-controlled interlayer exchange coupling is
perfectly reversible
on thermal cycling within the given temperature range.  %

\begin{figure}[htbp]
\centering
\includegraphics[width=0.9\columnwidth]{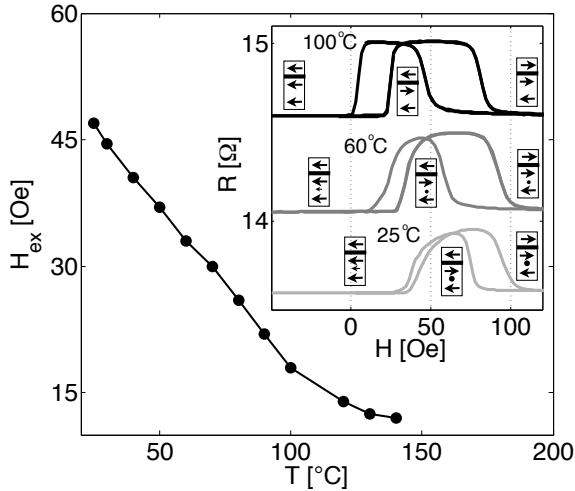}\centering
\caption{Interlayer exchange field H$_{\mathrm{ex}}$ versus
temperature $T$. The composition of the complete stack is
Si/SiO$_{2}$/Ni$_{80}$Fe$_{20}$ 3/ Mn$_{80}$Ir$_{20}$
15/Co$_{90}$Fe$_{10}$ 2/ Ni$_{70}$Cu$_{30}$ 30/Co$_{90}$Fe$_{10}$
2/Ni$_{80}$Fe$_{20}$10/Co$_{90}$Fe$_{10}$ 2 /Cu 7/Co$_{90}$Fe$_{10}$
4/Ni$_{80}$Fe$_{20}$ 3/Mn$_{80}$Ir$_{20}$ 15/Ta 5 [nm]. Inset:
Current-in-plane GMR at T=25, 60 and $100^{\circ}$C.}
\label{fig:Hex_vs_T}
\end{figure}

We further demonstrate an exchange-biased magnetic tri-layer of the
generic composition
AF/F0/f/F2, where the spacer separating the outer
ferromagnetic layers (F) is a low-Curie temperature diluted
ferromagnetic alloy (f) and one of the
F0 layers is exchange-pinned
by an antiferromagnet (AF). In addition to the tri-layer a Cu spacer
and a reference layer, pinned by an AF, have been added on top of
the stack in order to measure the current-in-plane giant
magnetoresistance (GMR). The specific stack composition chosen was
Si/SiO$_{2}$/NiFe 3/MnIr 15/CoFe 2/ Ni$_{70}$Cu$_{30}$ 30/CoFe
2/NiFe 10/CoFe 2 /Cu 7/CoFe 4/NiFe 3/MnIr 15/Ta 5 [nm]. The sample
was deposited at room temperature in a magnetic field of 350 Oe,
then annealed at 300$^{\circ}$C for 20 minutes, and field cooled to
RT in $\sim800$ Oe. The NiCu spacer was co-sputtered while rotating
the substrate holder, such that the final concentration was 70$\%$
Ni and 30$\%$ Cu having the $T_{C}$ suitably above RT. Fig.
\ref{fig:Hex_vs_T} shows how the \emph{interlayer} exchange field
H$_{\mathrm{ex}}$ of this sample varies with temperature. H$_{ex}$
shown in the main panel of Fig. \ref{fig:Hex_vs_T} is defined as the
mid point switching field of the soft
F2-layer ($\sim18$, $\sim32$, and $\sim47$ Oe for $100^{\circ}$C, $60^{\circ}$C, and $25^{\circ}$C, respectively; see inset),
which reflects the strength of the interlayer exchange coupling
through the spacer undergoing a ferromagnetic-paramagnetic
transition in this temperature range.
To explain why this is so, we need to consider the difference in effective magnetic thickness between the top and bottom pinned ferromagnetic layers. The effective magnetic thickness for the bottom pinned CoFe/NiCu/CoFe/NiFe/CoFe layers is approximately three times larger than for the top pinned CoFe/NiFe. From the inset to Fig 4, the temperature variation of the exchange pinning for the top pinned CoFe/NiFe is ~20 Oe or 0.3 Oe/K. If we were to assume that the bottom pinned CoFe/NiCu/CoFe/NiFe/CoFe layers are coupled and reverse as one layer, and that the variation in exchange field is caused solely by the weakening pinning at the bottom MnIr interface, then we would expect an exchange field three times smaller than for the top pinned CoFe/NiFe. With a three times smaller exchange field at RT the expected temperature variation would be 7 Oe or 0.1 Oe/K, which clearly is much lower than the observed change of 25 Oe (from ~45 Oe to ~20 Oe) and therefore the measured de-pinning of the switching layer is predominantly due to a softening of the exchange spring.

We have separately measured the strength of the exchange pinning at the bottom MnIr surface. For CoFe ferromagnetic layers 2-4 nm thick, the pinning strength at RT is 500 Oe or more. At 130 C, at which the spacer is paramagnetic and fully decoupled from the underlying MnIr/CoFe bilayer, the pinning strength is still above 100 Oe. We therefore conclude that the dominating effect in question is the weakening exchange spring in the spacer.
This demonstrates the principle of the thermionic spin-valve
proposed, where the P to AP switching is controlled by temperature.
The AF-pinned implementation of the spin-thermionic valve presented
should be highly relevant for
application. %

As is obvious from the above analysis the dependence of the
magnetization direction on temperature  allows  electrical
manipulations of it by Joule heating  with an applied current
flowing through the stack. In the next section we find  connection
between the magnetization direction and the current-voltage
characteristics (IVC) of such a spin-thermionic valve.


\section{Thermoelectric manipulation of the magnetization direction.}
\subsection{Current-voltage characteristics of the stack under
Joule heating.}

 If the stack is Joule heated  by a current $J$ its
temperature $T(V)$ is determined by  the heat-balance condition
\begin{equation}
JV=Q(T), \hspace{0.2cm}J  =V/R(\Theta),
 \label{heat}
\end{equation}
and Eq.~(\ref{theta2}), which determines the temperature dependence
of $\Theta(T(V))$. Here $Q(T)$ is the heat flux from the stack and
$R(\Theta)$ is the stack resistance. In the vicinity of the Curie
temperature $T_c^{(1)}$ Eq.~(\ref{theta2}) can be re-written as
\begin{eqnarray}
\Theta = \Biggl\{\begin{array}{l} \pm \pi, \hspace{2.4 cm} T \geq T_c^{(1)}
 \\
D_0 \frac{T_c^{(1)} }{T_c^{(1)}-T}\sin\Theta,\hspace{0.5 cm} T <
T_c^{(1)}\,,\end{array}
 \label{theta2modif}
\end{eqnarray}
(here $D_0$ is defined in Eq.(\ref{Torient})).

Equations~(\ref{heat}) and (\ref{theta2modif}) define the
current-voltage characteristics (IVC) of the stack, $J=G(\Theta(V))V$, $G=R^{-1}$, in a parametric
form
 which can be re-written as
\begin{eqnarray}
J=\sqrt{Q(T_c^{(1)})}\sqrt{G(\theta)(1-{\bar
D}\frac{\sin{\theta}}{\theta})}
\nonumber \\
V=\sqrt{Q(T_c^{(1)})}\sqrt{R(\theta)(1-{\bar
D}\frac{\sin{\theta}}{\theta})}\,. \label{IVCparametric}
\end{eqnarray}
The parameter $\theta$ is defined in the interval $-\pi \leq \theta
\leq \pi$,
$${\bar D}=D_0 \frac{T}{Q}\frac{dQ}{dT}\bigr|_{T=T_c^{(1)}}\approx
D_0\,,$$
and in order to derive Eq.~(\ref{IVCparametric}) we used the
expansion $Q(T)=Q(T_c^{(1)}) +Q'_T(T_c^{(1)})(T-T_c^{(1)})$ [$Q'_T
\equiv dQ/dT$].

\begin{figure}
 \centerline{\includegraphics[width=0.85\columnwidth]{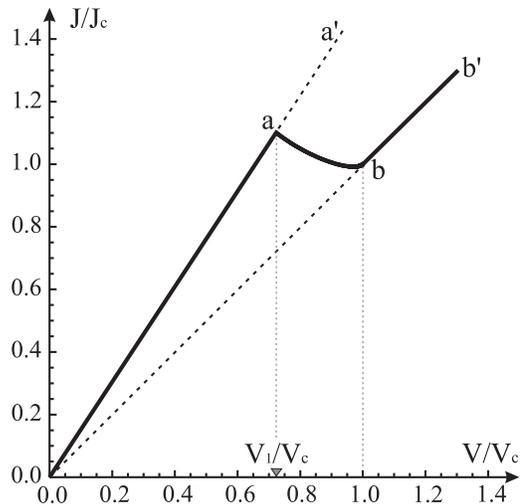}}
\vspace{0 mm} \caption{ Current-voltage characteristics (IVC) of the
magnetic stack of Fig.~\ref{noflip} calculated for
$R(\Theta)=R_{+}-R_{-} \cos{\Theta}$, $R_-/R_+ =0.2$, $D_0=0.2$;
$J_c=V_c/R(\pi)$. The branches  $0-a$ and $b-b'$ of the IVC
correspond to parallel and antiparallel orientations of the stack
magnetization, respectively (the parts $a-a'$ and $0-b$ are
unstable); the branch $a-b$ corresponds to the inhomogeneous
magnetization distribution shown in Fig.~\ref{changeorient}.}
 \label{cvc}
\end{figure}

It follows from  Section \ref{equilibrium} that the stack resistance is
$R(0)$ in the entire temperature range $T(V) < T_c^{\rm(or)}$ and
$R(\pi)$ in the range $T(V)>T_c^{(1)}$. This implies that the
 IVC branches $J=G(0)V$
 and $J=G(\pi)V$ are linear for,
respectively,
\begin{equation}
 V< V_{1}=\sqrt{R(0)Q(T_c^{\rm(or)})}
 \label{V1}
\end{equation}
 ($0-a$ in
Fig.~\ref{cvc}) and
\begin{equation}
 V > V_{c}=\sqrt{R(\pi)Q(T_c^{(1)})}
 \label{Vc}
\end{equation}
($b-b'$
in Fig.~\ref{cvc}). If $V_1 \leq V \leq V_c$ the stack temperature
is $T_c^{\rm(or)} \leq T(V) \leq T_c^{(1)}$, and the direction of
the magnetization in region 2
changes  with
a change of $V$; hence
the IVC is non-linear there. Below we find the conditions under
which this branch of the IVC has a negative differential
conductance.

Differentiating Eq.~(\ref{IVCparametric}) with respect to $V$ one
finds
\begin{equation}
\frac{dJ}{dV}=R \left(\Theta \right) \frac{[G\left(\theta) (1-{\bar
D}\sin{\theta}/\theta\right)]'}{[R(\theta) (1-{\bar
D}\sin{\theta}/\theta)]'}\Bigr|_{\theta=\Theta(V)}
 \label{diffG(T)}
\end{equation}
where
$[\ldots]'$ means the derivative of the bracketed quantity with
respect to the angle $\theta$, and $\Theta(V)$ is found from the
second equation in Eq.~(\ref{IVCparametric}). From this result it
follows
that the differential conductance $G_{d}(V) \equiv d J/dV$ is negative if
$$\frac{d}{d\Theta}\frac{(1-{\bar D}\sin{\Theta}/\Theta)}{ R(\Theta)}<0\,.$$
For a stack resistance of the form
\begin{equation}
R(\Theta)= R_{+}\left(1 -r \cos{\Theta}\right)
\label{Rtheta}
\end{equation}
where
\begin{equation}
r=\frac{R_-}{R_{+}}; \hspace{0.5cm} R_{\pm}=\frac{R(\pi)\pm R(0)}{2}
\label{r}
\end{equation}
one finds that the differential conductance $dJ/dV < 0$  if
\begin{equation}
D_0 <\frac{3r}{1+2r}
\label{criteria}
\end{equation}
Hence the IVC of the stack is
N-shaped as shown in Fig.~\ref{cvc}.

 We note here that the modulus of
 the negative differential conductance may be
  large  even in the case that   the magnetoresistance  is small. Using Eq.(15)
  at $r\ll 1$ one finds the differential conductance $G_{diff}$ as
 \begin{equation}
G_{diff}\equiv  \frac{d J}{dV}= -R^{-1}(0)\frac{1-D_0/3r}{1+D_0/3r}
 \label{diffres1}
 \end{equation}
 which is negative provided $D_0 < 3r$, the modulus of $G_{diff}$  being   of the order of $R^{-1}(0)$.

 Here and below we consider the case that the electric current flowing through the sample is lower than the torque critical current and hence the torque effect is absent \cite{torquecurrent}

  \begin{figure}
 \centerline{\includegraphics[width=0.85\columnwidth]{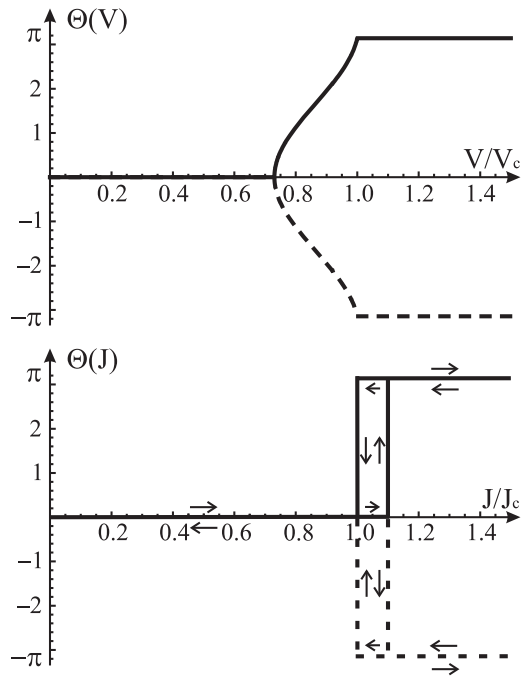}}
  \caption{The angle $\Theta$, which describes the tilt of the direction of
  the magnetization in layer 2 with respect to that in layer 1 (see Fig.~\ref{changeorient}),
  as a function of voltage in the voltage-biased regime (top) and
  current in the current-biased regime (bottom). Both curves were
  calculated for $R(\Theta)=R_{+}-R_{-} \cos{\Theta}$, $R_-/R_+ =0.2$,
$D_0=0.2$; $J_c=V_c/R(\pi)$.}
 \label{theta-V}
 \end{figure}
As the IVC curve $J(V)$ is N-shaped  the thermoelectrical  manipulation of the relative
 orientation of  layers 0 and 2 may be of two different types depending on the ratio between the resistance
 of the stack and resistance of  the circuit in which it is incorporated

In the voltage-bias regime
 which corresponds to the case that the resistance of the
stack is much larger than resistance of the rest of the circuit, the voltage drop across the stack
preserves the given value which is approximately equal to the bias voltage and hence there is only
one value of the current (one point on the  IVC) $J=J_{bias}$ corresponding to the bias voltage $V_{bias}$
(see Fig.\ref{cvc})
In this case the relative orientation of the
magnetization of layers 0 and 2 can be changed smoothly from being
parallel to anti-parallel by varying the bias voltage through the
interval $V_1 \leq  V_{\rm bias}\leq V_c$. This corresponds to
moving along the $a-b$ branch of the IVC. The dependence of the
magnetization direction $\Theta$ on the voltage drop across the
stack is shown in Fig.~\ref{theta-V}.

In the current-bias regime, on the other hand,  which corresponds to the case that the resistance of the stack
is much smaller than the resistance of the circuit,  the current in the circuit $J$ is kept at a given value which is mainly determined by the bias voltage and the circuit resistance (being nearly independent of the stack resistance). As this takes place,  the voltage drop across  the stack $V$ differs from the bias voltage $V_{bias}$, being  determined by the equation $J(V)=J$. As the IVC is N-shaped, the stack
may now be in a bistable state: if the current is between points $a$ and $b$  there are three possible values of the voltage drop across the stack  at one fixed value of the current (see Fig. \ref{cvc}).  The states of the stack with the lowest and the highest voltages across it are stable while the state of the stack with  the middle value of the voltage drop is unstable. Therefore, a change of the current results in a
hysteresis loop as shown in Fig.~\ref{theta-V}: an increase of the
current along the $0-a'$ branch of the IVC leaves the magnetization
directions in the stack parallel ($\Theta=0$) up to point $a$,
where the voltage drop $V$ across the stack jumps to the right branch $b-b'$, the jump
being accompanied by a fast switching of the stack magnetization
from the parallel to the antiparallel orientation ($\Theta=\pm
\pi$). A decrease of the current along the $b'-0 $ IVC branch keeps
the stack magnetization antiparallel up to point $b$, where the
voltage jumps to the left $0-a'$  branch of the IVC  and the magnetization of the stack
 comes back to the parallel orientation  ($\Theta=0$).

In the next Section we will show that this scenario for a
thermal-electrical manipulation of the magnetization direction is
valid for small values of the inductance in the electrical circuit.
If the inductance exceeds some critical value the above steady state
solution becomes unstable and spontaneous oscillations appear in the
values of the current, voltage drop across the stack, temperature,
and direction of the magnetization.
\begin{figure}[htbp]
\centering \includegraphics[width=0.5\columnwidth]{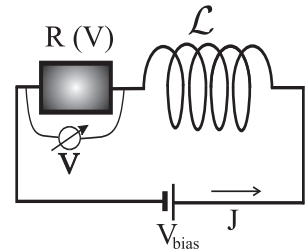}
\caption{ Equivalent circuit for a Joule-heated magnetic stack of
the type shown in Fig.~\ref{noflip}.
  A  resistance $R(V)=J(t)/V(t)$, biased by a fixed DC voltage $ V_{\rm bias}$, is
  connected in series with an
  inductance ${\cal L}$; $V(t)$ is the voltage drop over the
  stack and $J(t)$ is the total current.} \label{circuit}
\end{figure}


\subsection{ Self-excited electrical, thermal and directional
magnetic oscillations. \label{selfoscillations}}
\subsubsection{Current perpendicular to layer planes (CPP)\label{CPP}}
Consider now a situation in the bias voltage regime
where the magnetic
stack under investigation is connected in series with an inductance
${\cal L}$ and biased by a DC voltage $V_{\rm bias}$, as described
by the equivalent circuit in Fig.~\ref{circuit}.
The thermal and
electrical processes in this system are governed by the set of
equations
\begin{equation}
C_V\frac{dT}{dt}=J^2 R(\Theta)-Q(T); \;\ \;
{\cal L}\frac{dJ}{dt} +JR(\Theta)= V_{\rm bias}\,, 
 \label{evolution}
\end{equation}
where $C_V$ is the heat capacity. The relaxation of the magnetic
moment to its thermodynamically equilibrium direction is assumed to be the fastest process in the problem,  which implies that  the magnetization direction corresponds to the equilibrium state of the stack at the given temperature $T(t)$. In other words,
the  tilt angle,
$\Theta=\Theta(T(t))$, adiabatically follows the time-evolution of the temperature and hence its temperature dependence is given by Eq.~(\ref{theta2}).

 \begin{figure}
  \centerline{\includegraphics[width=0.75\columnwidth]{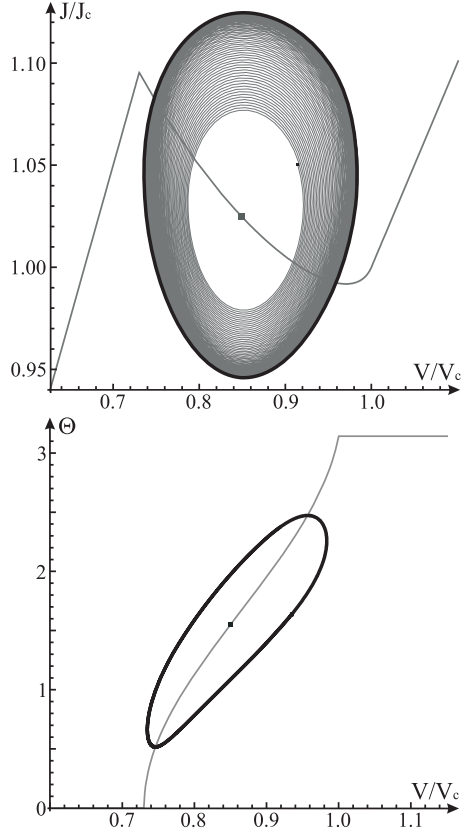}}
  \caption{Spontaneous oscillations of the current
$J(t)$
  and the voltage drop $ V(t)$ over the stack calculated for
  $R_-/R_+=0.2$,  $D_0=0.2$ and
$({\cal L} -{\cal L}_{\rm cr})/{\cal L}_{\rm cr}=0.013$;
$J_c=V_c/R(\pi)$. $J(t)$ and $ V(t)$ develop
from the initial state  towards
  the limit cycle (thick solid line) along which they execute a periodic motion. The thin line is the stationary IVC of the stack.
 The bottom figure shows the limit cycle  along which $\Theta(t)$ and $ V(t)$ execute a periodic motion.}
   \label{cvclimmiddle}
  \end{figure}
A time dependent variation of the temperature is accompanied by a
variation of the magnetization angle $\Theta(T(t))$  and hence by
a change in the voltage drop across the stack via the dependence of
the magneto-resistance on this angle, $R=R(\Theta)$.

The system of equations Eq.(\ref{evolution})  has one time-independent  solution ($\bar{T}(V_{bias}), \hspace{.1cm} \bar{J}(V_{bias}) $)
which is determined by the equations
\begin{equation}\label{SteadySolution}
J^2R\left(\Theta (T)\right)=Q(T),\hspace{1cm} J R(\Theta (T))=V_{bias}
\end{equation}
 This solution is identical to the solution of Eqs.(\ref{theta2},\ref{heat}) that determines the N-shaped IVC shown in Fig.\ref{cvc} with a change $J\rightarrow \bar{J}$ and  $V\rightarrow V_{bias}$.

In order to investigate the stability of this time-independent solution we write the temperature, current and the angle as a sum of two terms,

\begin{eqnarray}
T= \bar{T}(V_{bias})+T_1(t); \nonumber \\
J=\bar{J}(V_{bias})+J_1(t); \nonumber \\ 
\Theta=\bar{\Theta}(V_{bias}) +\theta_1(t),
\label{eqlinear}
\end{eqnarray}

where $T_1$, $J_1$ and $\theta_1$ each is a
small correction. Inserting Eq.(\ref{eqlinear}) into Eq.(\ref{evolution}) and Eq.~(\ref{theta2}) one easily finds that the time-independent solution Eq.(\ref{SteadySolution})
is always stable at any value of the inductance ${\cal L}$ if the bias voltage $V_{bias}$ corresponds to a branch of the IVC with a positive differential resistance (branches 0-a and b-b' in Fig.\ref{cvc}).
If the bias voltage $V_{bias}$ corresponds to the branch with a negative differential resistance ($V_1<V_{bias}<V_c$, see Fig.\ref{cvc}) the solution of the set of linearized equations is  $T_1=T_1^{(0)} \exp\{\gamma t\}$, $J_1=J_1^{(0)}\exp\{\gamma t\}$ and $\theta_1=\theta_1^{(0)}\exp\{\gamma t\}$ where $T_1^{(0)} $, $J_1^{(0)}$ and $\theta_1^{(0)}$ are any initial values close to the steady-state of the system, and
\begin{equation}
\gamma=\frac{\bar{R}}{2{\cal L}}\Bigl( \frac{{\cal L}-{\cal L}_{c}}{{\cal L}_{c}}\pm \sqrt{\left(\frac{{\cal L}-{\cal L}_{c}}{{\cal L}_{c}}\right)^2-4\frac{|R_{d}|}{\bar{R}}\frac{{\cal L}}{{\cal L}_{c}}}\Bigr)
 \label{gamma}
\end{equation}
where
\begin{equation}
{\cal L}_c=\frac{C_V}{|d(GQ)/dT|}\Bigl|_{T=T(V)}
 \label{criticalinductance}
\end{equation}
and $R_{d}=d V/dJ$, $\bar{R} = R(\bar{\Theta})$ is the differential resistance.

As is seen from Eq.(\ref{gamma})
the steady-state solution Eq.(\ref{SteadySolution}) is stable only if the inductance ${\cal L}\leq {\cal L}_c$; if the inductance exceeds the critical value Eq.(\ref{criticalinductance}) the system looses its stability and a limit cycle
appears in the plane $(J,T)$ (see, e.g., \cite{Hilborn}). This corresponds to the appearance of self-excited, non-linear and periodic temporal oscillations of the
temperature $T=T(t)$ and the current $J=J(t)$, which are accompanied by oscillations of the voltage drop across the stack ${\tilde V}(t)=J(t)$ and the the magnetization direction $\Theta(t) = \Theta(T(t))$. For the case that $({\cal L}-{\cal L}_c )/{\cal L}_c \ll 1$ the system executes nearly harmonic oscillations around the steady state (see Eq.(\ref{eqlinear})) with the frequency $\omega =\rm Im \gamma ({\cal L}={\cal L}_c)$, that is the temperature $T$, the current $J$, the magnetization direction $\Theta$ and the voltage drop across the stack $V(t)=R(\Theta(t))J(t)$ execute a periodic motion with the frequency
\begin{equation}
\omega = \frac{\sqrt{\bar{R} R_d}}{{\cal L}_c}
\label{frquency}
\end{equation}

With a further increase of the inductance the size of the limit cycle grows, the amplitude of the oscillations increases and the oscillations become  anharmonic,
the period of the oscillations therewith decreases  with an increase of the inductance ${\cal L}$.

In order to investigate the time evolution of the voltage drop across the stack and
the current   in more details it is convenient
to introduce an auxiliary
voltage drop ${\tilde V}(t)$  and a current $ J_0(t)$ related to
each other through Eqs.~(\ref{heat}) and (\ref{theta2modif}). Hence
we define
\begin{equation}
 {\tilde V}(t)= \sqrt{R(T(t))Q(T(t))) };\;\ \; J_0={\tilde
 V}(t)/R(T(t))\,,
 \label{VtildeJ}
\end{equation}
where $R(T)=R(\Theta(T))$. Comparing these expressions with
Eq.~(\ref{heat}) one sees  that at any moment $t$
Eq.~(\ref{VtildeJ}) gives the stationary IVC of the stack, $J_0= J_0({\tilde V})$, defined by Eq.~(\ref{IVCparametric})
(changing $J \rightarrow J_0$ and $V \rightarrow {\tilde
V}$), see Fig.~(\ref{cvc}).

Differentiating ${\tilde V}(t)$ with respect to $t$  and using
Eqs.~(\ref{evolution}) and (\ref{VtildeJ}) one finds that the
dynamical evolution of the system is governed by the equations
\begin{eqnarray}
\tau_0 \frac{d {\tilde V}}{d t}=\frac{J^2 -J_0^2({\tilde
V})}{2J_0({\tilde V})}
\nonumber \\
{\cal L} \frac{d  J}{d t}+ \frac{J{\tilde V}}{J_0({\tilde
V})}=V_{\rm bias}
 \label{evolutionJV}
\end{eqnarray}
where $$\tau_0=  \frac{C_V }{(QR)_T'}\Bigr|_{T=T({\tilde V})}.$$ As
follows from the second equation in Eq.~(\ref{VtildeJ}), at  any
moment $t$ the voltage  drop over the stack $V(t)=R(T(t))J(t)$ is
coupled with ${\tilde V}(t)$ by the following relation:
$$V= \frac{J}{J_0({\tilde V})}{\tilde V}.$$

\begin{figure}
  \centerline{\includegraphics[width=0.85\columnwidth]{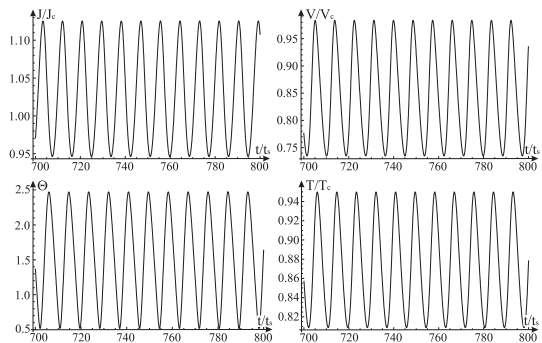}}
  \caption{Spontaneous oscillations of the current $J(t)$,
   the voltage drop $V(t)$, the magnetization direction
  angle $\Theta(t)$, and the temperature $T(t)$ corresponding to motion along the
limit cycle shown in Fig.\ref{cvclimmiddle}.  Calculation parameters are
  $R_-/R_+=0.2$,  $D_0=0.2$ and
$({\cal L} -{\cal L}_{\rm cr})/{\cal L}_{\rm cr}=0.3\times 10^{-4}$;
$J_c=V_c/R(\pi)$. }
   \label{limthetasmall}
  \end{figure}

The coupled equations (\ref{evolution}) have only one  steady-state
solution  $J=J_0(V_{bias})$ where $J_0(V)  $ is the IVC shown in Fig.\ref{cvc}
(see   Eqs.~(\ref{VtildeJ}). However, in
the interval $V_1 \leq V_{\rm bias} \leq V_c$ this solution is
unstable with respect to small perturbations if ${\cal L} > {\cal
L}_{\rm cr}$. As a result periodic oscillations of the
current $J(t)$ and $ {\tilde V}(t)$ appear spontaneously, with
$J(t)$, $ {\tilde V}(t)$ eventually reaching a limit cycle. The limiting cycle  in the
$J$-$V$ plane
is shown in Fig.~\ref{cvclimmiddle}.
 The stack temperature
 $T=T(t)$, the magnetization direction $\Theta(t)= \Theta(T(t))$,
 follow these electrical oscillations adiabatically according to the
 relations $Q(T(t))= {\tilde V}(t)J_0(t)$ (here $J_0(t)\equiv J_0({\tilde V}(t))$) and $\Theta(t)=\Theta(T(t))$
 (see Eq.~(\ref{theta2})) as shown in  Fig.~\ref{limthetasmall}.

 \begin{figure}
  \centerline{\includegraphics[width=0.85\columnwidth]{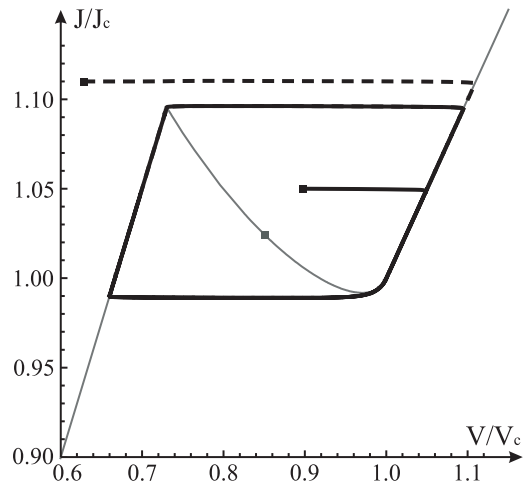}}
  \caption{Spontaneous oscillations of the current
$J(t)$
  and the voltage drop ${\tilde V}(t)$  calculated for
  $R_-/R_+=0.2$,  $D_0=0.2$ and
$({\cal L} -{\cal L}_{\rm cr})/{\cal L}_{\rm cr}=535$;
$J_c=V_c/R(\pi)$. The time development of $J(t)$ and ${\tilde V}(t)$
follows one or the other of the dashed lines towards
  the limit cycle (thick solid line) depending on whether
  the initial state is inside or outside the limit cycle.
  The bottom figure shows how the current oscillations develop if
  the initial state is inside the limit cycle. The stationary IVC of the stack is shown as a thin solid line.}
   \label{limIVbig}
  \end{figure}

 \begin{figure}
  \centerline{\includegraphics[width=0.85\columnwidth]{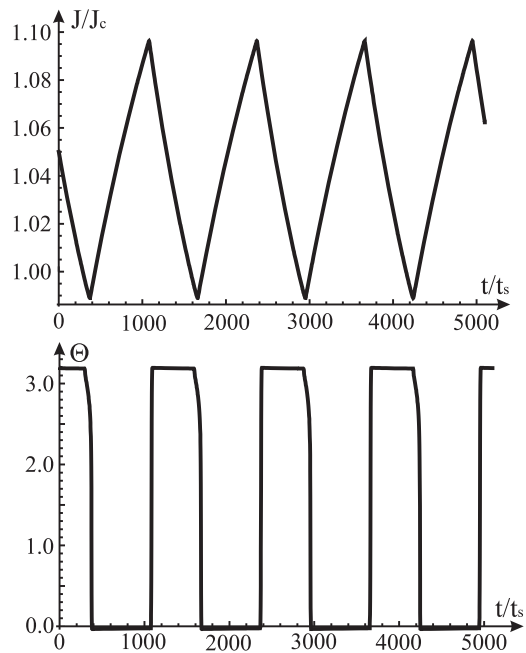}}
  \caption{Spontaneous oscillations of the magnetization direction
  angle $\Theta(t)$
   calculated for
  $R_-/R_+=0.2$,  $D_0=0.2$ and
$({\cal L} -{\cal L}_{\rm cr})/{\cal L}_{\rm cr}=535$;
$J_c=V_c/R(\pi)$.}
   \label{limthetabig}
  \end{figure}

The character of the oscillations changes  drastically in the limit
 ${\cal L}\gg {\cal L}_{\rm cr}$. In this case the current and the voltage slowly
move along the branches $0 - a$  and $b - b^{'}$
 of the IVC at the rate ${\dot J}/J \approx
{R_+/\cal L}$,
  quickly switching between these branches
at the points $a$  and $b$  with the rate $\sim
1/\tau_0$ (see Fig.~\ref{limIVbig}). Therefore, in this case the stack
periodically switches between the parallel and antiparallel magnetic
states (see Fig.~\ref{limthetabig}).

\subsubsection{Current in the layer planes (CIP).}
If the electric current flows in the plane of the layers (CIP)
of the stack  the torque effect is insufficient or absent \cite{Slonczewski,Ralph}  while the magneto-thermal-electric oscillations under consideration may take place. In this case the total current flowing through
the cross-section of the layers may be presented as
\begin{equation}
 J_{CIP}= \left(R^{-1}(\Theta) + R^{-1}_{0}\right)V
 \label{CIP}
\end{equation}
where   $R(\Theta)$ and $R_{0}$ are the magneto-resistance
and the angle-independent resistance of the stack in the CIP set of the experiment.

In  a CIP configuration the stack is Joule heated by  both the angle-dependent and the angle-independent currents and hence Eq.(\ref{heat}) should be re-written as follows:
\begin{equation}
J_{CIP}V=Q(T), \hspace{0.2cm}J  =V/R_{eff}(\Theta),
 \label{heatCIP}
\end{equation}
where
\begin{equation}
R_{eff}(\Theta)=\frac{R(\Theta) R_{0}}{R(\Theta) + R_{0}}
 \label{Reff}
\end{equation}

Using Eq.(\ref{diffG(T)}) and Eq.(\ref{CIP}) one finds that the presence of the angle-independent current in the stack modifies the condition of the negative differential conductance $dJ_{CIP}/dV$: it is negative if
\begin{equation}
\bar{D} <  \frac{3r}{(1+2r)R_0+ (1-r)^2(1-4r)R_+}\Bigl[ R_0 -(1-r)^2 R_+ \Bigr]
 \label{}
\end{equation}
As is seen from here, an IVC  with a negative differential resistance
is possible if $R_0 >(1-r)^2 R_+ $  (see Eq.[\ref{r}] for definitions of $R_{\pm}$ and $r$).

The time evolution of the system  is described by the set
of equations Eq.(\ref{evolution}) in which one needs to change $J\rightarrow J_{CIP}$ and $R(\Theta) \rightarrow R_{eff}(\Theta)$. Therefore, under this change, the temporal evolution of the system in a CIP configuration is the same as when  the  current  flows perpendicular to the stack layers: if the bias voltage corresponds to the negative differential conductance $dJ_{CIP}/dV <0$ and the inductance exceeds the critical value
\begin{equation}
{\cal L}_c=\frac{C_V}{|d(G_{eff}Q)/dT|}\Bigl|_{T=T(V)}
 \label{criticalinductanceEff}
\end{equation}
where $G_{eff}=R_{eff}^{-1}$, self-excited oscillations of the current $J_{CIV}$, voltage drop over the stack
 $V$, the temperature $T$  and the angle $\Theta(T(V))$ arise in the system, the maximal frequency of which being
\begin{equation}
\omega = \frac{\sqrt{|dV/dJ_{CIP}|R_{eff}(T(V))}}{{\cal L}_{cr}}\Bigl|_{V=V_{bias}}
\label{frequencyEff}
\end{equation}
if ${(\cal L}-{\cal L}_{\rm cr})/{\cal L}_{\rm cr}\ll1$

Below we present estimations of the critical inductance and the oscillation frequency which are valid  for  both the above mentioned CPP and CIP configurations of the experiment.

 Using equations Eq.(\ref{SteadySolution}) and Eq.(\ref{criticalinductance}) one may estimate the order of magnitude  of the critical inductance and the oscillation frequency as ${\cal L}_c\approx T c_v/j^2d$ and $\omega \approx \rho j^2/T c_v$ where $c_v$ is the heat capacity per unit volume, $\rho$ is the  resistivity, and $d$ is a characteristic size of the stack. For point contact devices with   typical values of $d \sim 10^{-6} \div 10^{-5} \rm cm$, $c_v \sim 1\; \mathrm{J/cm}^3$K, $\rho\sim 10^{-5} $ $\Omega$cm,  $j\sim 10^8 \;\mathrm{ A/cm^2}$ and
assuming that cooling of the device can provide the sample temperature $T\approx T_c^{(1)}\sim 10^2 \rm K$ one finds the characteristic values of the critical inductance and the oscillation frequency as
${\cal L}_{\rm cr}\approx 10^{-8}\div 10^{-7}\rm H$
and
$\omega \approx 1 {\rm GHz}$

\section{Conclusions.} The experimental implementation of the new
principle proposed in this paper for the electrical manipulation of
nanomagnetic conductors by means of a controlled Joule heating of a
point contact appears to be quite feasible. This conclusion is
supported both by theoretical considerations and preliminary
experimental results, as discussed in the main body of the paper.
Hence we expect the new spin-thermo-electronic oscillators that we
propose to be realizable in the laboratory. We envision F0/f/F2 valves
where two strongly ferromagnetic regions ($T_c \sim 1000$ K) are
connected through a weakly ferromagnetic spacer ($T_c \ll 1000$ K).
The Curie temperature of the spacer would be variable on the scale
of room temperature, chosen during fabrication to optimize the
device performance. For example, doping Ni-Fe with $\sim 10 \%$ of
Mo brings the $T_c$ from $\sim 1000$ K to 300-400 K. Alternatively,
alloying Ni with Cu yields a spacer with a $T_c$ just above Room
Temperature (at RT or below RT, if needed). If a sufficient current
density is created in the nano-tri-layer to raise the temperature to
just above the $T_c$ of the spacer, the magnetic subsystem
undergoes a transition from the F0/f/F2 state to an F0/N/F2 state, the
latter being similar to conventional spin-valves (N for nonmagnetic,
paramagnetic in this case). Such a transition should result in a
large resistance change, of the same magnitude as the ``giant
magnetoresistance" (GMR) for the particular material composition of
the valve.

Local heating (up to 1000 K over 10-50~ nm) can readily be produced
using, e.g.,
point contacts in the thermal
regime, with very modest global circuit currents and essentially no
global heating \cite{Yanson}.
Heat is known to propagate through nm-sized objects on the ns
time scale, which can be scaled with size to the sub-ns regime. When
voltage-biased to generate a temperature near $T_c(f)$, such a F0/f/F2
device would oscillate between the two magnetic states, resulting in
current oscillations of a  frequency that can be tuned by means of
connecting a variable inductance in series with the device.
Spin rotation frequencies  may be tuned from the GHz-range  down
to quasi-DC (or DC as soon as the inductance is smaller than the
critical value).
For F0/f/F2 structures geometrically designed in the style of the
spin-flop free layer of today's magnetoresistive random access memory (MRAM), the dipolar coupling between
the two strongly ferromagnetic layers would make the anti-parallel
state ($F0\uparrow /N/F2 \downarrow$
) the magnetic ground state above
$T_c(f)$. The thermal transition in the f-layer would then drive a
full 180-degree spin-flop of the valve. The proposed spin-thermo-electronic valve can be implemented in CPP as well as CIP geometry, which should make it possible to achieve MR signals of ~10.

In conclusion, we have shown that Joule heating of the magnetic
stack sketched in Fig.~\ref{noflip} allows the relative orientation
of the magnetization of the two ferromagnetic layers 0 and 2 to be
electrically manipulated. Based on this principle, we have proposed
a novel spin-thermo-electronic oscillator concept and discussed how
it can be implemented experimentally.

{\it Acknowledgement.} Financial support from the Swedish VR and
SSF, the European Commission (FP7-ICT-2007-C; proj no 225955 STELE)
and the Korean WCU programme funded by MEST through KOSEF
(R31-2008-000-10057-0) is gratefully acknowledged.


\begin{thebibliography}{99}
\bibitem{torquecurrent} Typical densities of critical currents needed for  the torque effect in point-contact devices are $10^8\div 10^9 \rm A/cm^2$  for the current  perpendicular to the layers (CPP) \cite{Ralph}
\end{thebibliography}
\end{document}